\documentclass[pra,twocolumn,showpacs,floatfix]{revtex4}
\usepackage[applemac]{inputenc}
\usepackage{amsmath,amssymb}
\usepackage{mathrsfs}
\usepackage{graphicx}

\newcommand{\ave}[1]{\left \langle {#1} \right \rangle}

\newcommand{\ket}[1]{\left \lvert {#1} \right \rangle}
\newcommand{\kB}{k_{\mathrm{B}}}
\newcommand{\V}{\mathrm{V}}

\newcommand{\var}{{\mathrm{var}}}
\newcommand{\fano}{\mathscr{F}}
\newcommand{\spectral}{\mathscr{S}}

\begin{document}

\title{Rate-equation approach to atomic-laser light statistics.}

\author{Laurent Chusseau}

\email{chusseau@univ-montp2.fr}

\homepage{http://www.opto.univ-montp2.fr/~chusseau}

\affiliation{Centre d'Électronique et de Micro-optoélectronique de
Montpellier, UMR 5507 CNRS, Université Montpellier II, F34095
Montpellier, France}

\author{Jacques Arnaud}

\email{arnaudj2@wanadoo.fr}

\affiliation{Mas Liron, F30440 Saint Martial, France}

\author{Fabrice Philippe}
\altaffiliation[Also at ]{MIAp, Université Paul Valéry, F34199 
Montpellier, France}

\email{Fabrice.Philippe@univ-montp3.fr}

\affiliation{Laboratoire d'Informatique de Robotique et de
Microélectronique de Montpellier, UMR 5506 CNRS, 161 Rue Ada, F34392
Montpellier, France}

\date{\today}

\begin{abstract}
	We consider three- and four-level atomic lasers that are either
	incoherently (unidirectionally) or coherently (bidirectionally)
	pumped, the single-mode cavity being resonant with the laser
	transition.  The intra-cavity Fano factor and the photo-current
	spectral density are evaluated on the basis of rate equations. 
	According to that approach, fluctuations are caused by jumps in
	active \emph{and} detecting atoms.  The algebra is considerably
	simpler than the one required by Quantum-Optics treatments. 
	Whenever a comparison can be made, the expressions obtained
	coincide.  The conditions under which the output light exhibits
	sub-Poissonian statistics are considered in detail.  Analytical
	results, based on linearization, are verified by comparison with
	Monte Carlo simulations.  An essentially exhaustive investigation
	of sub-Poissonian light generation by three- and four-level atoms
	lasers has been performed.  Only special forms were reported
	earlier.
	
\end{abstract}


\maketitle

\section{Introduction}

Interest in the statistics of light emitted by atomic lasers has been
recently revived as a result of the realization of micro-lasers
\cite{an:PRL94}.  The main purpose of this paper is to show that
expressions derived from rate equations coincide with those previously
derived from Quantum Optics.  This is so even when the generated light
exhibits sub-Poissonian statistics.  An introduction to that method as
it pertains to sub-Poissonian light generation can be found in
tutorial papers \cite{arnaud:OQE95,arnaud:OQE01}.  Let us recall that
light is called sub-Poissonian when the variance of the number of
photo-detection events over some large time duration is less than the
average number of events.  Equivalently, one may say that the spectral
density of the photo-current is less than the shot-noise level at
small Fourier (or baseband) frequencies.  Analytical expressions are
obtained from rate equations in a straightforward manner.  Namely, the
expression for the internal cavity statistics of many 4-level atoms
with a negligible spontaneous decay previously given in Eq.~(4) by
\citet{ritsch:PRA91} is recovered exactly (see Eq.~\eqref{newS} of the
present paper).  Similarly, 3-levels atoms expressions obtained by
\citet{khazanov:PRA90} are recovered.  But coherently-pumped 3-level
atoms lasers were apparently not treated earlier.  If the upper and
lower decay times of 4-level atoms tend to zero, the laser is
equivalent to a 2-levels atoms laser with Poissonian pump
\cite{chusseau:2levels}.  In that limit, the expressions reported in
\citet{arnaud:IEEJ90} in \citeyear{arnaud:IEEJ90} are recovered.  When
the above approximations are not applicable, the presently reported
expressions appear to be new.

\begin{figure*}
    \begin{tabular}{c@{\hspace{1.5cm}} c@{\hspace{1.5cm}} c}
        (a) & (b) & (c)  \\
        \includegraphics[scale=0.6]{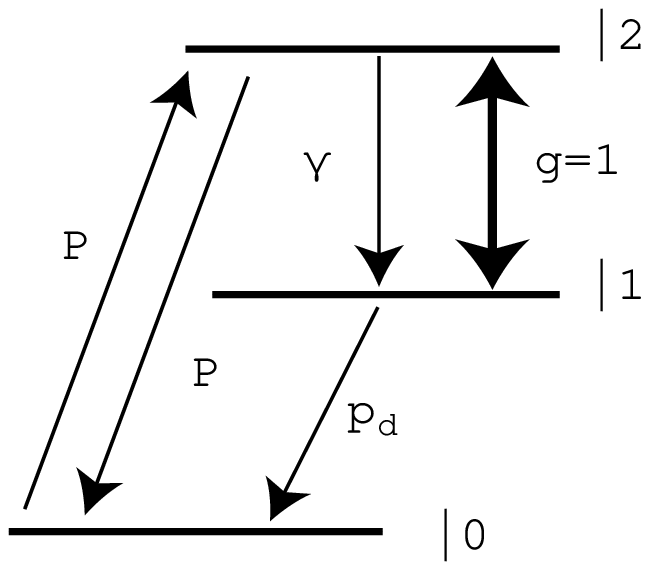} & 
        \includegraphics[scale=0.6]{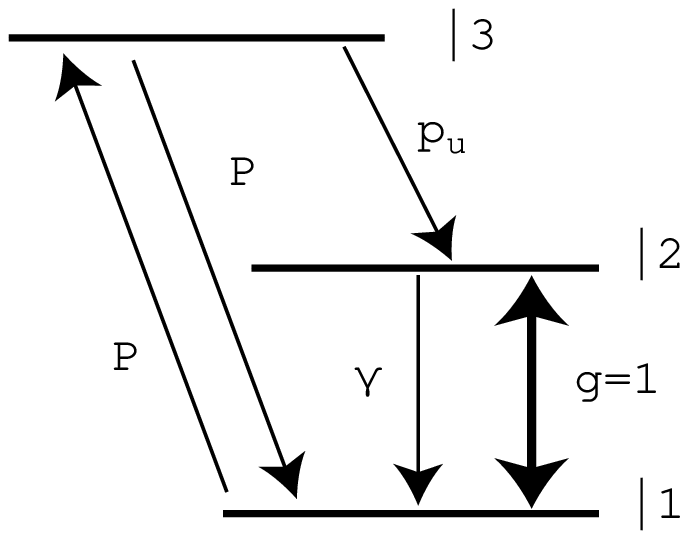} & 
        \includegraphics[scale=0.6]{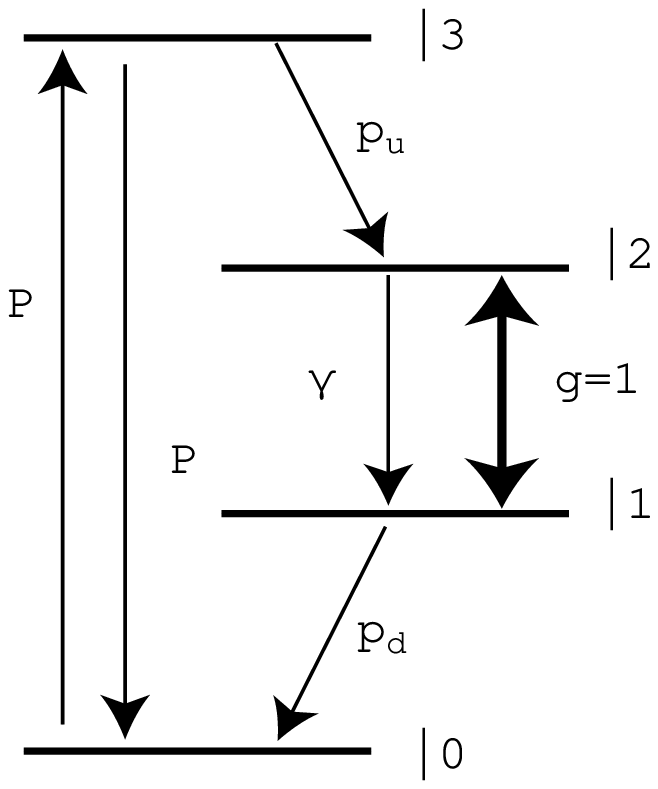}  \\
    \end{tabular}
	\caption{Level schemes for atomic lasers.  (a) $\Lambda$--type
	3-level laser, (b) $\V$--type 3-level laser, (c)
	4-level atoms laser.  For incoherent pumping $\ell = 0$.  For
	coherent pumping $\ell = 1$.}
    \label{fig:0}
\end{figure*}

The active medium is a collection of $N$ identical atoms, with the
levels labeled $\ket{0}$, $\ket{1}$, $\ket{2}$, and $\ket{3}$ in
increasing order of energy (see Fig.~\ref{fig:0}).  Level separations
are supposed to be large compared with $\kB T$, where $T$ denotes the
optical cavity temperature and $\kB$ the Boltzmann constant, so that
thermally-induced transitions are negligible.  Note that the
3-level atoms schemes in Fig.~\ref{fig:0}(a) and \ref{fig:0}(b)
are \emph{not} special cases of the 4-level scheme in
Fig.~\ref{fig:0}(c).  They are treated separately in appendices. 
Levels $\ket{1}$ and $\ket{2}$ are resonant with the field of a
single-mode optical cavity.  The active medium is supposed to be
strongly homogeneously broadened so that atomic polarizations may be
adiabatically eliminated, an approximation applicable also to YAG
lasers, CO$_{2}$ lasers and semiconductor lasers.  Such lasers may
operate in the steady-state regime, that is, pulsation and chaos do
not normally occur.

The probability per unit time that an electronic transition from level
$\ket{1}$ to level $\ket{2}$ occurs is taken as equal to $m$, and the
probability of an electronic transition from $\ket{2}$ to $\ket{1}$ as
$m+1$, where $m$ denotes the number of photons in the cavity (the
qualification ``per unit time'' is henceforth omitted for the sake of
brevity).  This amounts to selecting a time unit whose typical value
depends on the gain medium.  Spontaneous decay from level $\ket{2}$ to
level $\ket{1}$ is allowed with probability $\gamma$.  This decay may
be either non-radiative or involve radiation into other
electromagnetic modes, besides the one of interest.  Photons are
absorbed with probability $\alpha m$, where $\alpha$ denotes a
constant, the absorbing atoms residing most of the time in their
ground state.  These absorbing atoms model the transmission of light
through mirrors with subsequent absorption by a detector.  Provided
detection is linear and reflectionless, it is immaterial whether
absorption occurs inside or outside the optical cavity.  For
simplicity, internal absorption is neglected.

``Incoherent'' pumping promotes electrons from level $\ket{0}$ to
level $\ket{3}$ with probability $P$.  When transitions from $\ket{0}$
to $\ket{3}$ and from $\ket{3}$ to $\ket{0}$ are both allowed with
equal probabilities, the pumping process is called ``coherent'',
following an accepted terminology.  We find it convenient to denote by
$\ell P$ the $\ket{3} \to \ket{0}$ transition probability, with $\ell
= 0$ for incoherent pumping and $\ell = 1$ for coherent pumping. 
Coherent pumping is physically realized by submitting the atoms to
strong optical fields nearly resonant with the $\ket{0} \to \ket{3}$
transition.  This pumping field may possibly originate from
frequency-filtered thermal radiation and be highly multimode.  Levels
$\ket{0}$ and $\ket{3}$ need not be sharp.  Instead, they may consist
of narrow bands for improved coupling to broad-band pumps.  One-way
incoherent pumping would be appropriate to describe laser-diode pumps. 
In semiconductors, however, working levels spread into conduction and
valence bands, so that the present model would not be adequate.  Laser
diodes have been treated previously on the basis of rate equations in
\cite{arnaud:OQE01}.  Spontaneous decay from level $\ket{3}$ to the
upper working level $\ket{2}$ occurs with probability $p_{u}$, and
spontaneous decay from the lower working level $\ket{1}$ to the ground
level with probability $p_{d}$, where ``u'' stands for ``up'' and
``d'' for down.  The relevant probabilities are schematized in
Fig.~\ref{fig:0}.

One of the best-known laser-noise theory is probably that of
\citet{scully:PR67}.  Incoherent pumping is modeled by independent
injection of 2-level atoms in the optical cavity.  This model leads to
a photo-count statistics which is, at best, Poissonian.  More
recently, \citet{khazanov:PRA90,ralph:PRA91,ritsch:PRA91} considered
the situation in which the pumping-levels populations may fluctuate. 
At first, it would seem that this may only increase the noise.  It
turns out, however, that population fluctuations are correlated in
such a way that the output light fluctuations may be sub-Poissonian. 
It is difficult to pin point a simple intuitive explanation.  It has
been observed, however, that when lasers are pumped through a cascade
of intermediate levels, pumping tends to be regular
\cite{ritsch:PRA92,briegel:PRA96}, a situation somewhat similar to
laser-diodes high-resistance driving conditions.

Other means of generating sub-Poissonian light have been considered. 
\citet{golubev:JETP84} were the first in \citeyear{golubev:JETP84} to
point out that lasers with non-fluctuating pumps should emit
sub-Poissonian light.  This conclusion has been verified
experimentally by \citet{yamamoto:PRL87} with the help of laser diodes
driven by high-impedance electrical sources.  The
\citeauthor{scully:PR67} model has been generalized to account for
regular atom injection \cite{levien:PRA93,zhu:PRA93}. 
\citet{kolobov:PRA93} made the interesting observation that the
photodetection rate spectral density may be below the shot-noise level
at \emph{non-zero} Fourier frequencies in the case of
\emph{Poissonian} pumps.  However, the photodetection rate spectral
density remains at the shot-noise level at \emph{zero} frequency. 
Accordingly, such lasers do not generate sub-Poissonian light in the
sense defined earlier.  It has been shown that 3-level lasers with
coherent decay to the ground state \cite{haake:PRL93,golubev:TMP96}
generate sub-Poissonian light, and further that Raman lasers may
generate sub-Poissonian light \cite{ritch:EUL92}.  We will not
consider here these more exotic configurations.  A review is in
\cite{davidovitch:RMP96}.

Rate equations treat the number of photons in the cavity as a
classical random function of time.  The light field is quantized as a
result of matter quantization and conservation of energy, but not
directly.  Rate equations should be distinguished from semi-classical
theories in which the optical field is driven by atomic dipole
expectation values.  The theory employed in this paper rests instead
on the consideration of transition probabilities, as in the
\citet{loudon} treatment of optical amplifier noise, for example. 
Every absorption event reacts on the number of light quanta in the
optical cavity.  Semi-classical theories are unable to explain
sub-Poissonian light statistics because the light generation process
and the light detection processes are considered separately.  The
expressions obtained from rate equations are found to coincide with
Quantum Optics results when the number of atoms is large and
transitions other than those relating to the atom-cavity interaction
are incoherent.  The laser is treated as a birth-death Markov process
(see Sec.~\ref{sec.MC}).  A Monte Carlo simulation gives the evolution
of the number $m$ of photons in the cavity from which the Fano factor
$\fano =\var(m)/\ave{m}$ is obtained.  On the other hand,
the instants $t_{k}$ when photons are being absorbed provide the
spectral density of the photo-current, whose normalized value
$\spectral$ is unity for Poisson processes.  The normalized spectrum
is denoted $\spectral(\Omega)$.  The Fourier angular frequency
$\Omega$ is called for short: ``frequency''.

When both the number of atoms and the pumping level increase, the
computing time becomes prohibitively large because of the
exponentially growing number of events to process.  Analytical results
are obtained by applying the weak-noise approximation.  We first give
in Sec.~\ref{subsec.SS} the steady-state photon number.  The
photo-current spectral density at zero frequency is evaluated in
Sec.~\ref{subsec.0f}, and the photo-current spectrum in
Sec.~\ref{subsec.non0f}.  The method of evaluation of the intra-cavity
Fano factor is in Sec.~\ref{subsec.Fano}.

\section{Monte Carlo method}
\label{sec.MC}

The rate-equation model of a $N$--atoms singlemode laser
straightforwardly leads to a master equation for the probability of
having $m$ photons stored in the cavity at time $t$
\cite{scully:PR67,loudon,arnaud:OQE01}.  Alternatively, the laser
evolution is modeled as a temporally homogeneous birth-death Markov
process.  In the steady-sate regime, rate of change and equilibrium
probabilities of having $m$ and $m+1$ photons within the cavity are
linked via a detailed balancing condition.  This is a favorable
condition for a Monte Carlo simulation \cite{landau} because every
laser microstate belongs to a Markov chain and will thus occurs
proportionally to its equilibrium probability when the number of step
increases to $\infty$.

For the $\V$--type lasers considered in the present section (see
Fig.~\ref{fig:0}(b)), rates of change $W_{j}$ are ascribed to the
different kinds of events as given in Table~\ref{tbl:1}.  For example,
the probability that an atom jumps from level $\ket{1}$ to level
$\ket{3}$ during the elementary time interval $[t,t+\delta t]$ is
$W_{2} \delta t$, where $\delta t$ is chosen small enough that this
probability be much less than unity.  Because atoms are coupled to one
another only through the field, $W_{2}$ is proportional to the number
$n_{1}$ of atoms in $\ket{1}$ at time $t$, and thus $W_{2} = P n_{1}$,
where the constant $P$ is proportional to the pump strength.  If a
jump does occur, $n_{1}$ is reduced by 1 while the number $n_{3}$ of
atoms in level $\ket{3}$ is incremented by 1.  If the initial value of
$n_{3}$ is $N$ the event does not occur.  Similar observations apply
to the other jump probabilities.  Notice that the coherent emission
rate $W_{4}$ is proportional to $m+1$, following the Einstein
prescription.  This ensures that laser emission re-starts if
extinction occurs.  A key feature that distinguishes the present
formulation from other rate-equation methods is that absorption of
photons by the detector is included in the system description. 
Because detection is supposed to be linear, such events are taken to
occur with a rate $W_{1} = \alpha m$, where $\alpha$ expresses
detector absorption.

\begin{table}
	\caption{Elementary events in $\V$--type 3-level lasers (see
	Fig.~\ref{fig:0}(b)) and corresponding rate of change $W_{j}$.}
    \label{tbl:1}
    \begin{ruledtabular}
		\begin{tabular}{rcl}
			Event & Transition & Rate \\
			photon absorption & --- & $W_{1} = \alpha m$ \\
			pump absorption & $\ket{1} \to \ket{3}$ & $W_{2} = P
			n_{1}$ \\
			pump emission & $\ket{3} \to \ket{1}$ & $W_{3} = \ell P
			n_{3}$ \\
			coherent emission & $\ket{2} \to \ket{1}$ & $W_{4} =
			\left( m + 1 \right) n_{2}$ \\
			coherent absorption & $\ket{1} \to \ket{2}$ & $W_{5} = m
			n_{1}$ \\
			spontaneous decay & $\ket{2} \to \ket{1}$ & $W_{6} =
			\gamma n_{2}$ \\
			upper decay & $\ket{3} \to \ket{2}$ & $W_{7} = n_{3}
			p_{u}$ \\
		\end{tabular}
    \end{ruledtabular}
\end{table}

An efficient algorithm has actually been employed
\cite{gillespie}: Given that an event of any kind occurred at
time $\tau_{k}$, the next-event time is
\begin{equation}
	\tau_{k+1} = \tau_{k} + \frac{1}{\sum_{j} W_{j}} \ln \left(
	\frac{1}{\mathfrak{r}} \right) ,
    \label{eq:31}
\end{equation}
where $\mathfrak{r}$ is a random number uniformly distributed in the
interval $[0,1]$.  The probability that the event is of kind $l$ is
equal to $W_{l} / \sum_{j} W_{j}$.  The Monte Carlo method was
implemented on a desk computer.  Only the total number of atoms in
each state needs to be tracked (For fermions the numerical procedure
is significantly more involved \cite{chusseau:OQE01}).  Within the
whole time set $\{\tau_{k}\}$, we are mostly interested in the subset
$\{t_{k}\}$ of photon-absorption events.  It is straightforward in
principle to evaluate the photo-detection noise.  We also record $m
\left( t_{k} \right)$ to evaluate the mean value and variance of $m$.

\begin{figure}
    \begin{tabular}{rl}
		(a) & \\
		& \includegraphics[width=7.5cm]{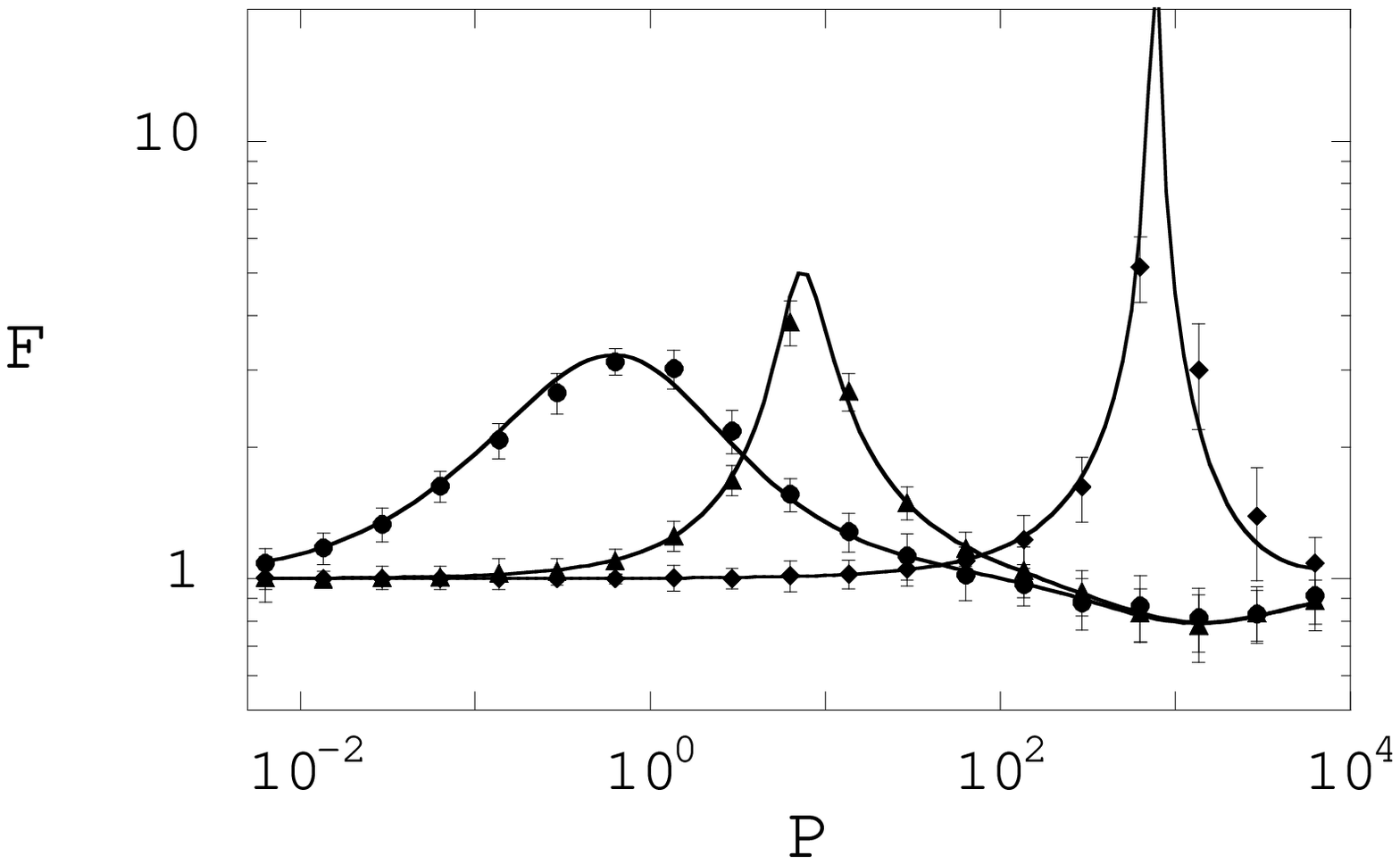} \\
		(b) & \\
		& \includegraphics[width=7.5cm]{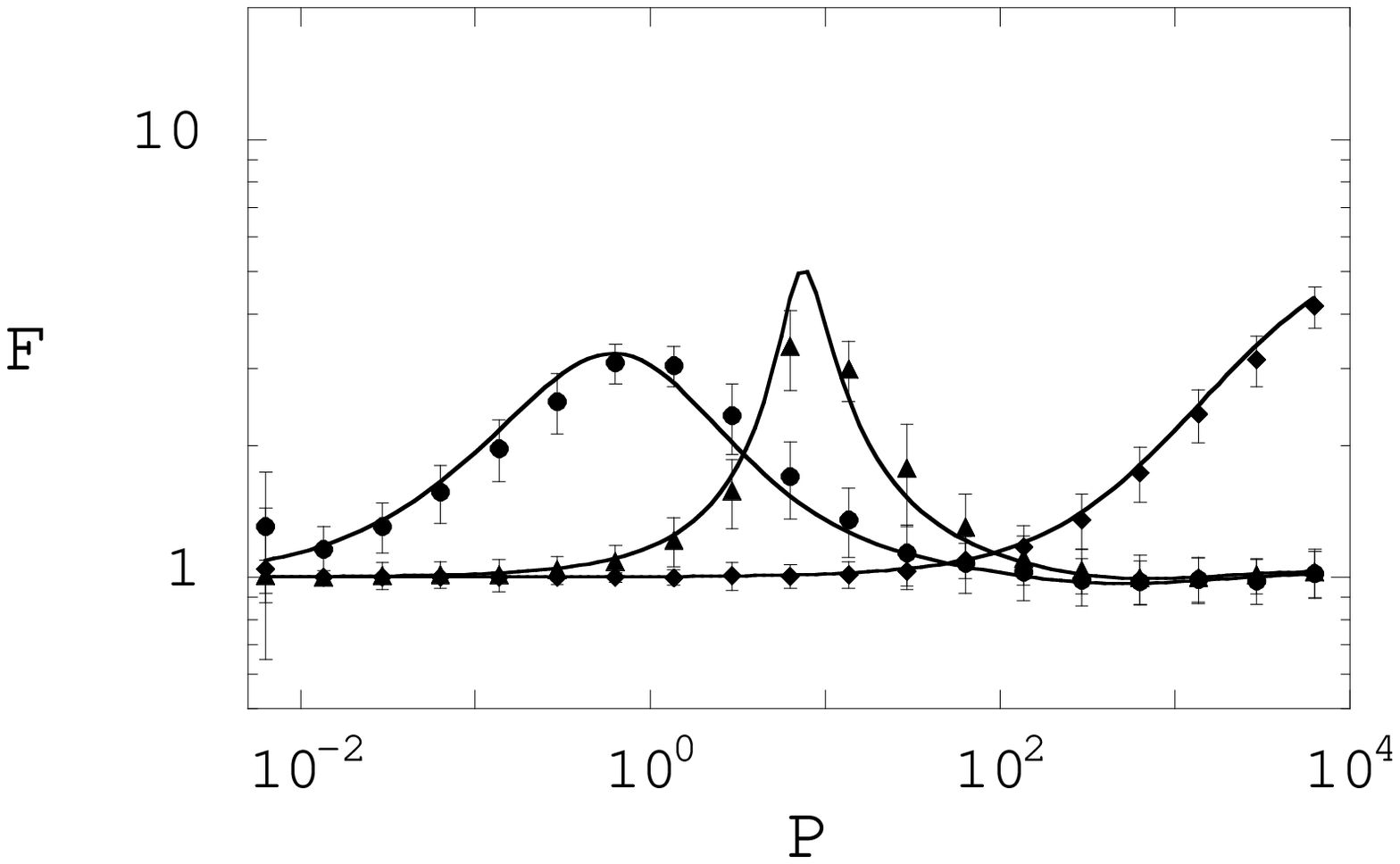} \\
    \end{tabular}
	\caption{Intra-cavity Fano factor for $\V$--type lasers as a
	function of the pumping rate $P$.  (a) incoherent pumping, (b)
	coherent pumping.  Error bars are the 95\,\% confidence level from
	a statistical treatment applied to ten Monte Carlo runs, each
	having duration $T_{m}=200$.  Plain lines are analytical (see
	Sec.~\ref{subsec.Fano}).  The parameters are: $N = 100$, $p _{u} =
	632$, $\alpha = 6.32$.  $\bullet$, $\gamma = 0$; $\blacktriangle$,
	$\gamma = 6.32$; $\blacklozenge$, $\gamma = 632$.}
    \label{FanoMC}
\end{figure}

The intra-cavity Fano factor $\fano$ is represented in
Fig.~\ref{FanoMC}.  Both coherent (Fig.~\ref{FanoMC}(a)) and
incoherent (Fig.~\ref{FanoMC}(b)) pumping are considered.  The
analytical results in Sec.~\ref{subsec.Fano}, based on linearization
agree well with the simulation.  Notice that $\fano$ is below unity
within some pumping range.  There is good agreement with previous
Quantum-Optics results \cite{koganov:PRA00}.

\begin{figure}
    \begin{tabular}{rl}
		(a) & \\
		& \includegraphics[width=7.5cm]{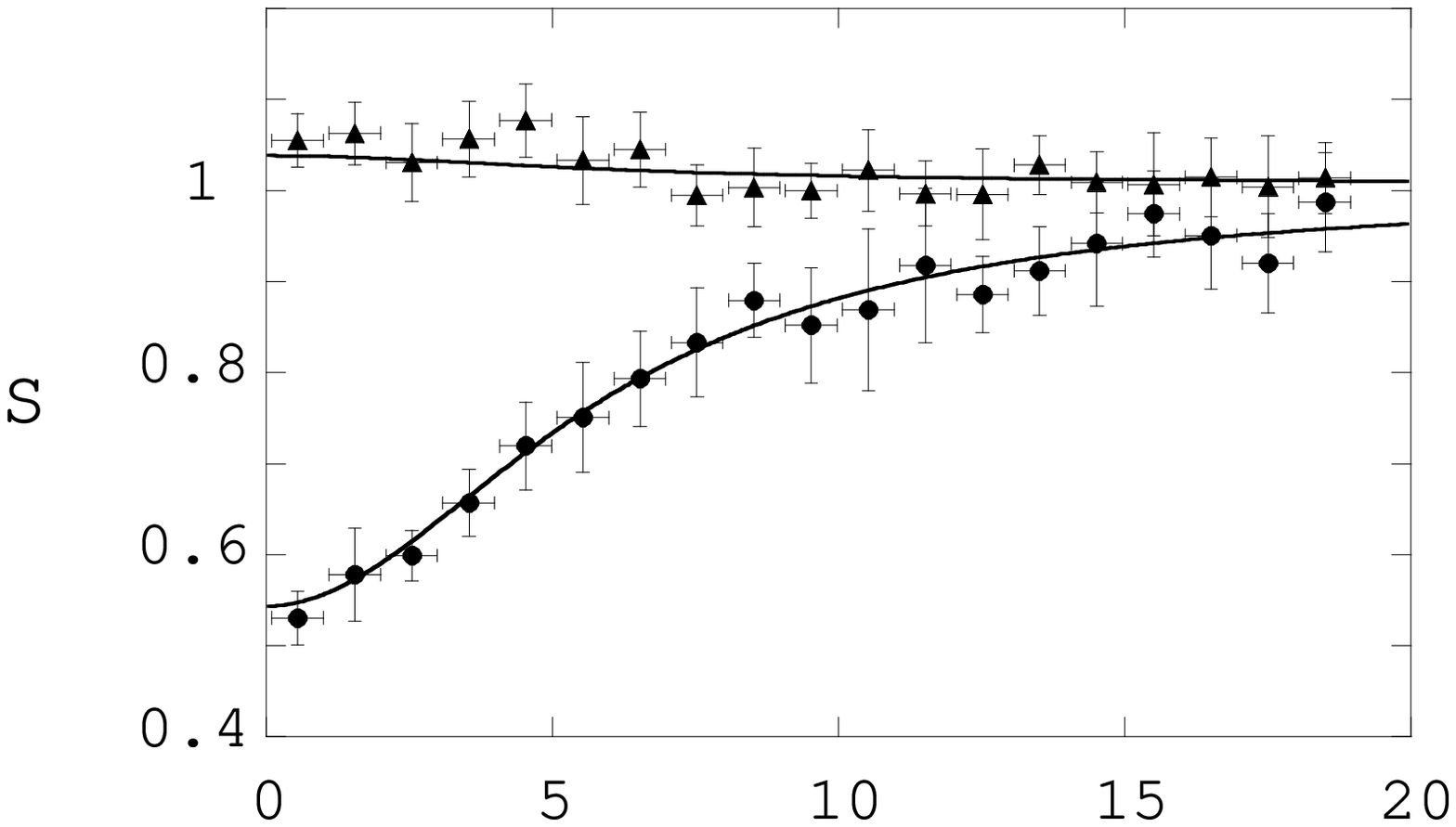} \\
		(b) & \\
		& \includegraphics[width=7.5cm]{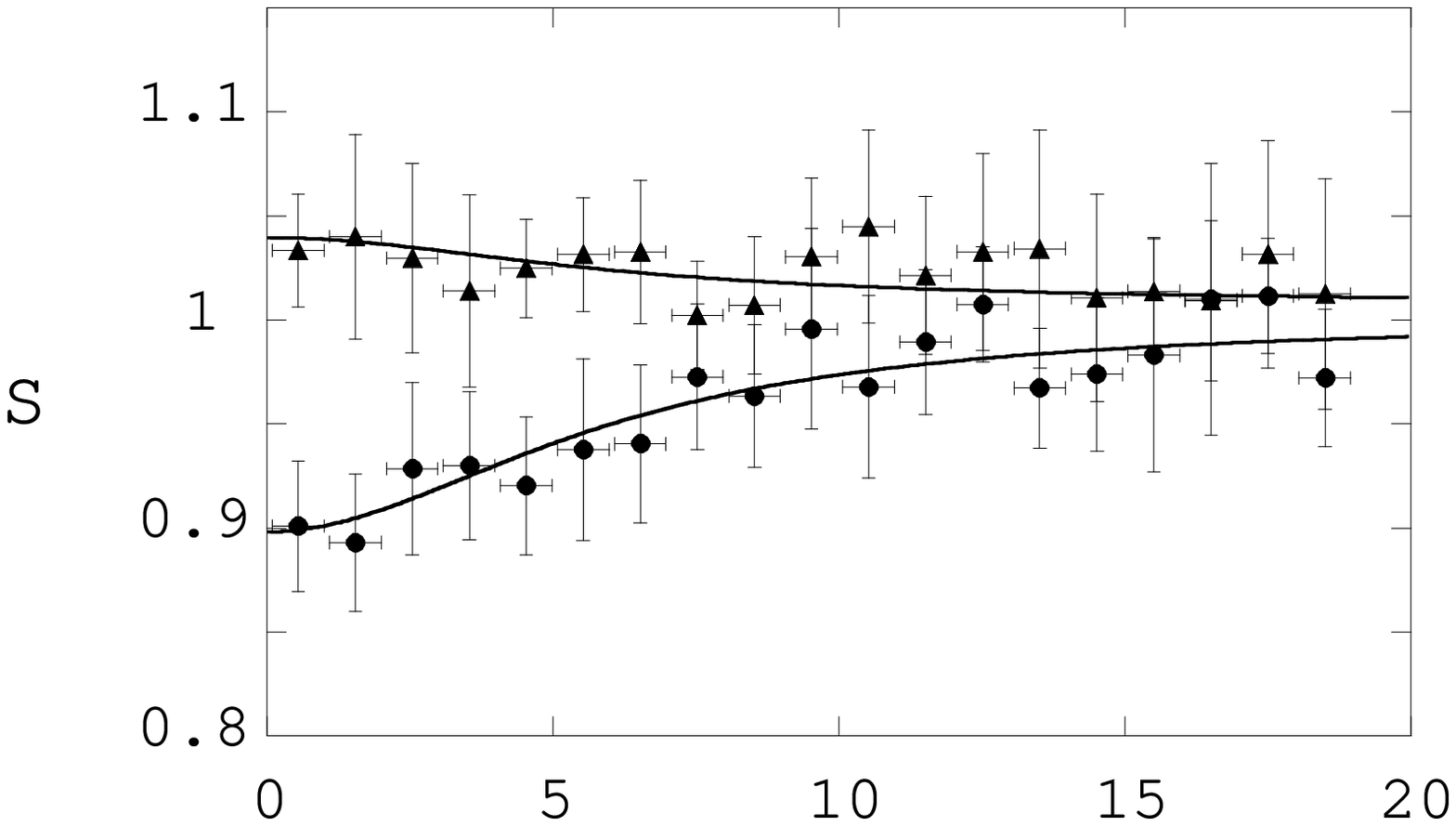} \\
    \end{tabular}
	\caption{Normalized photo-current spectral density $\spectral$ of
	$\V$--type lasers as a function of Fourier frequency $\Omega$. 
	Points with error bars are from Monte Carlo simulations with 150
	runs, each of duration $T_{m}=100$.  Plain lines are analytical
	(see Sec.~\ref{subsec.non0f}).  The parameters are: $N=100$,
	$p_{u}=632$, $\alpha = 6.32$.  (a) incoherent pumping $\bullet$,
	$\gamma = 0$ and $P=1265$; $\blacktriangle$, $\gamma =6.32$ and
	$P=20$.  (b) coherent pumping, $\bullet$, $\gamma = 0$ and
	$P=422$; $\blacktriangle$, $\gamma = 6.32$ and $P=20$.  }
    \label{SDQMC}
\end{figure}

The normalized spectral density $\spectral(\Omega)$ is represented in
Fig.~\ref{SDQMC} for incoherent (a) and coherent (b) pumps, and two
sets of parameter-values.  For each Monte Carlo run,
$\spectral(\Omega)$ is first evaluated from the $\{t_{k}\}$ list
\cite{chusseau:OQE01} and refined using a smoother power spectral
density estimator \cite{papoulis,press}.  Averaging over runs and
concatenating neighboring frequencies produce the final data together
with error bars at the 95\% confidence level.  There is fair agreement
between Monte Carlo simulations and analytical formulas to be
subsequently reported.  Both predict sub-Poissonian photo-current
statistics.  Even with one billion photon-absorption events, Monte
Carlo spectra exhibit large error bars.  An analytical method is to be
preferred when it exists.  On the other hand, Monte Carlo simulations
do not rely on linearization and provide a useful check.

\section{Four-levels atoms lasers}
\label{sec.Anal}

\subsection{Steady-state}
\label{subsec.SS}

Let $n_{j}$, $j = 0, 1, 2, 3$, denote the number of atoms in
state $j$, with
\begin{equation}
    n_{0}+n_{1}+n_{2}+n_{3}=N .
    \label{conservation}
\end{equation}
The transition rate from $\ket{1} \to \ket{2}$ (stimulated absorption)
is set equal to the number $m$ of photons in the cavity, a rule that
defines a time unit.  Pumping of an atom from level $\ket{0}$ to level
$\ket{3}$ occurs with rate $P$.  We allow for a transition rate $\ell
P$ that an atom in level $\ket{3}$ decays back to level $\ket{0}$. 
Atoms in state $\ket{3}$ spontaneously decay to state $\ket{2}$ with
rate $p_{u}$, while atoms in state $\ket{1}$ spontaneously decays to
state $\ket{0}$ with rate $p _{d}$.  These spontaneous decays are
labelled ``upper'' and ``lower'' decay, respectively.  Note that when
$1/p_{u}=0$ the population of level $\ket{3}$ vanishes and any
dependence on $\ell$ must therefore drop out.  Photons are absorbed
with rate $\alpha m$.  Spontaneous decay from the upper to the lower
working levels occurs with rate $\gamma$.  Altogether, there are eight
kinds of events that may occur in the course of time.

Let $\mathcal{J}$ denotes the net pumping rate, $\mathcal{R}$ the net
stimulated rate, $\mathcal{U}$ and $\mathcal{D}$ the upper and lower
decay rates, $\mathcal{S}$ the spontaneous decay rate from the upper
to the lower working levels, and $\mathcal{Q}$ the photon absorption
rate.  The steady-state conditions then read
\begin{subequations}
    \label{steadystate}
    \begin{align}
		\mathcal{J} &= \mathcal{U} = \mathcal{D} = \mathcal{R} +
		\mathcal{S} , \\
		\mathcal{Q} &= \mathcal{R} ,
    \end{align}
\end{subequations}
where
\begin{subequations}
    \label{rates}
    \begin{align}
		\mathcal{J} &= P n_{0} - \ell P n_{3} , & \mathcal{U} &=
		p_{u}n_{3} , \\
		\mathcal{R} &= (m+1) n_{2} - m n_{1} , & \mathcal{D} &=
		p_{d}n_{1} , \\
		\mathcal{S} &= \gamma n_{2} , & \mathcal{Q} &= \alpha m .
    \end{align}
\end{subequations}

Equations \eqref{conservation}, \eqref{steadystate} and \eqref{rates}
provide the steady-state atomic populations
$n_{i}$ and photon number $m$. In particular
\begin{equation}
    \label{squareroot}
	m = \frac{1}{2} \left( \mathscr{B} + \sqrt{ \mathscr{B}^2 + 4
	\mathscr{P} \mathscr{N}} \right) ,
\end{equation}
where
\begin{subequations}
    \label{averagephoton}
    \begin{align}
		\mathscr{N} &= \frac{N}{\alpha} ,\\
		\mathscr{P} &= \left[ \frac{1}{P} + \frac{2}{p_{d}} +
		\frac{1+\ell}{p_{u}} \right]^{-1} , \\
		\mathscr{B} &= \mathscr{P} \left[ \mathscr{N} - 1 +
		\frac{1}{p_{d}} \left( 1 + \gamma - \mathscr{N} \gamma \right)
		\right] - \gamma - 1 .
    \end{align}
\end{subequations}

For moderate pump powers above threshold, $m$ increases linearly with
$P$.  The intercept with the $m=0$ axis defines the threshold pump
power.  Mathematically, this amounts to first replacing in previous
equations $m+1$ by $m$, and then setting $m=0$.  The laser is found to
oscillate for some pump rate provided
\begin{equation}
    \label{shorthands}
	\gamma < \left( \mathscr{N} - 1 \right) {\left[ \frac{ \mathscr{N}
	+ 1}{p_{d}} + \frac{1+\ell}{p_{u}} \right]}^{-1} .
\end{equation}
The limiting case is obtained at infinite pump power
\begin{equation}
    \label{mmax}
	\lim_{P \to \infty}m = \frac{\left( \mathscr{N} - 1 \right) \left(
	p_{d} - \gamma \right)}{\left( 1 + \ell \right)
	\frac{p_{d}}{p_{u}} + 2} - \gamma .
\end{equation}

\begin{figure}
    \includegraphics[width=7.5cm]{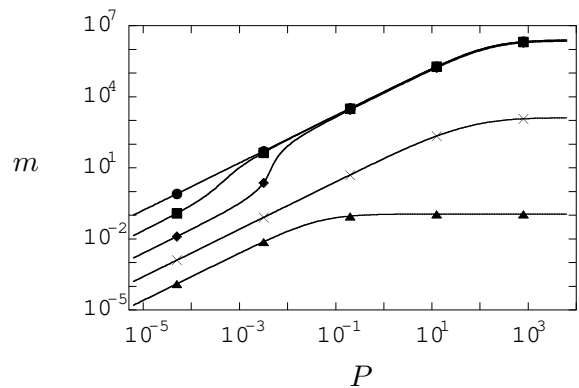} 
	\caption{Steady-state number of photons in 4-level atoms
	lasers as a function of $P/p_{d}$ for different values of
	the spontaneous decay rate $\gamma$.  $N=10^5$, $\alpha=6.32$,
	$p _{d}=632$.  For incoherent pumping $p _{u}=316$.  For
	coherent pumping $p_{u}=949$.  $\bullet$, $\gamma = 0$;
	$\blacksquare$, $\gamma = 6.32$; $\blacklozenge$, $\gamma = 63.2$;
	$\times$, $\gamma = 632$; $\blacktriangle$, $\gamma = 6325$.}
    \label{fig:m}
\end{figure}

Figure \ref{fig:m} shows how the steady-state photon number evolves as
a function of the pumping parameter $P$ for various values of the
spontaneous decay rate $\gamma $.  Parameters are similar to
\cite{koganov:PRA00} except $p_{u}$--values chosen here for a perfect
superimposition of the coherent and incoherent plots.  Moreover it
corresponds to optimum pump-noise suppression as derived in
Sec.~\ref{subsec.0f}.  At very high pump levels $m$ saturates like
Eq.~\eqref{mmax} because of ground-state population depletion.

\subsection{Zero-frequency noise}
\label{subsec.0f}

Our analytical results rest on a weak-noise approximation. 
Populations split into steady-state values and fluctuations.  For
example the instantaneous photon number $m$ is written as $\ave{m} +
\Delta m$, where $\ave{m}$ denotes the steady-state value.  Rates
split into steady-state values and fluctuations consisting of a
deterministic function of the population fluctuations and uncorrelated
Langevin ``forces''.  For example $\mathcal{J}$ split into $J =
\ave{\mathcal{J}}$ and $\Delta J$.  The latter is the sum of a
deterministic function of the population fluctuations and a Langevin
force $j(t)$ expressing the jump process randomness.  Thus
\begin{subequations}
    \label{fluctuations}
    \begin{align}
	\mathcal{J} &\equiv J + \Delta J , & \mathcal{U} &\equiv U +
	\Delta U , \\
	\mathcal{R} &\equiv R + \Delta R , & \mathcal{D} &\equiv D +
	\Delta D , \\
	\mathcal{S} &\equiv S + \Delta S , & \mathcal{Q} &\equiv Q +
	\Delta Q .
    \end{align}
\end{subequations}
where
\begin{subequations}
    \label{variations}
    \begin{align}
		\Delta J = &P \Delta n_{0} - \ell P \Delta n_{3} + j , \\
		\Delta R = &\left( m + 1 \right) \Delta n_{2} - m \Delta n_{1}
		\nonumber \\
		& + \left( n_{2} - n_{1} \right) \Delta m + r , \\
		\Delta S = & \gamma \Delta n_{2} + s , \\
		\Delta U = & p_{u} \Delta n_{3} + u ,\\
		\Delta D = & p_{d} \Delta n_{1} + d ,\\
		\Delta Q = & \alpha \Delta m + q .
    \end{align}
\end{subequations}
A first-order variation of the expressions in Eq.~\eqref{rates} has
been performed.

Conservation of the rates gives
\begin{subequations}
    \label{steadystateDelta}
    \begin{align}
		\Delta J &= \Delta U = \Delta D = \Delta R + \Delta S , \\
		\Delta Q &= \Delta R .
    \end{align}
\end{subequations}
Since the total number $N$ of atoms is constant we have
\begin{equation}
    \label{conservationDelta}
    \Delta n_{0} + \Delta n_{1} + \Delta n_{2} + \Delta n_{3} = 0 .
\end{equation}
Replacing atomic populations and photon numbers by their steady-state
values in Sec.~\ref{subsec.SS}, the above set of equations can be
solved.  In particular, $\Delta Q$ is a linear combination of the
Langevin forces
\begin{equation}
    \Delta Q = \sum_{z \in \{j,d,u,q,r,s\}}c_{z} z ,
    \label{DeltaQ}
\end{equation}
where the $c_{z}$ are real coefficients that depend on the parameters
$N$, $P$, $\ell$, $p_{u}$, $p_{d}$, $\gamma$ and $\alpha $.  The
detailed expressions, too lengthy to be given here in their general
form, are conveniently handled using symbolic calculations.

The normalized zero-frequency photo-current spectral density is of the
form
\begin{equation}
	\spectral = \frac{1}{\alpha m} \sum_{z \in \{j,d,u,r,s\}} c_{z}^2
	\sigma_{z} ,
    \label{SDfnull}
\end{equation}
where $\sigma_{z}$ denotes the spectral density value of the Langevin
noise source $z$, equal to average rates.
\begin{subequations}
    \label{SDlangevin}
    \begin{align}
		\sigma_{j} &= P n_{0} + \ell P n_{3} , & \sigma_{u} &= p_{u}
		n_{3} , \\
		\sigma_{r} &= \left( m + 1 \right) n_{2} + m n_{1} , &
		\sigma_{d} &= p _{d} n_{1} , \\
		\sigma_{s} &= \gamma n_{2} , & \sigma_{q} &= \alpha m .
    \end{align}
\end{subequations}
When these expressions are introduced in Eq.~\eqref{SDfnull} an analytical
expression of $\spectral$ is obtained. Three special cases are considered 
below, (a) $\gamma=0$ and $m$ large compared with unity, (b) 
$N \gg \alpha$, (c) $\gamma=0$ and $N \gg \alpha$.

(a) If spontaneous decay is negligible ($\gamma = 0$),
Eq.~\eqref{SDfnull} yields
\begin{multline}
	\label{SDgam0}
	\spectral = 1
	+ \frac{2}{{\left( \mathscr{N}-1 \right) }^2}
	+ \frac{8 \mathscr{P}^2}{p_{d}^2}
	+ \frac{\left( 6 - 4 \mathscr{N} \right) \mathscr{P} }{\left(
	\mathscr{N} - 1 \right) p_{d}} \\
	+ \frac{2 \left( 1 + \ell \right) \mathscr{P}^2}{p_{u}^2}
	+ \frac{2 \mathscr{P} \left( 2 \mathscr{P} - p_{d}
	\right)}{p_{d} p_{u}} ,
\end{multline}
where $\mathscr{P}$ and $\mathscr{N}$ are defined in
Eq.~\eqref{averagephoton}.  The normalized spectral density is unity
at low and high pumping levels.  For some constant $\mathscr{N}$
value, $\spectral$ reaches its minimum valued
\begin{equation}
    \label{SDgam0min}
	\spectral_{min} = \frac{2 \mathscr{N} \left( \mathscr{N} - 1
	\right) + 11 + \ell \left( 4 \mathscr{N} \left( \mathscr{N} - 1
	\right) + 15 \right)}{2 \left( 3 + 4 \ell \right) {\left(
	\mathscr{N} - 1 \right)}^2} ,
\end{equation}
when 
\begin{align}
	\label{SDgam0mincond}
	\frac{P}{p_{u}} &= \frac{1}{1 + \ell} , & \frac{P}{p_{d}} &=
	\frac{\mathscr{N} \left( 1 + 2 \ell \right) - \left( 2 + 3 \ell
	\right)}{ \left( 1 + \ell \right) \left( 2 \mathscr{N} - 1
	\right)} .
\end{align}

(b) When $N\gg \alpha$, Eq.~\eqref{SDfnull} yields
\begin{multline}
	\label{SDxnull}
	\spectral = 1
	+ \frac{2 \gamma}{p_{d} - \gamma}
	- \frac{4 \mathscr{P} + 2 \gamma}{p_{d}}
	+ \frac{8 \mathscr{P} \left( \mathscr{P} + \gamma \right)}{p_{d}^2}
	- \frac{8 \mathscr{P}^2 \gamma }{p_{d}^3} \\
	+ \frac{2 \mathscr{P} \left( \gamma - p_{d} \right) \left( p_{d} - 2 
	\mathscr{P} \right)}{p_{d}^2 p_{u}}
	- \frac{2 \left( 1 + \ell \right) \mathscr{P}^2 \left( \gamma - p_{d} 
	\right)}{p_{d} p_{u}^2} .
\end{multline}

\begin{figure}
    \begin{tabular}{cc}
		(a) & (b) \\
		\includegraphics[width=4.1cm]{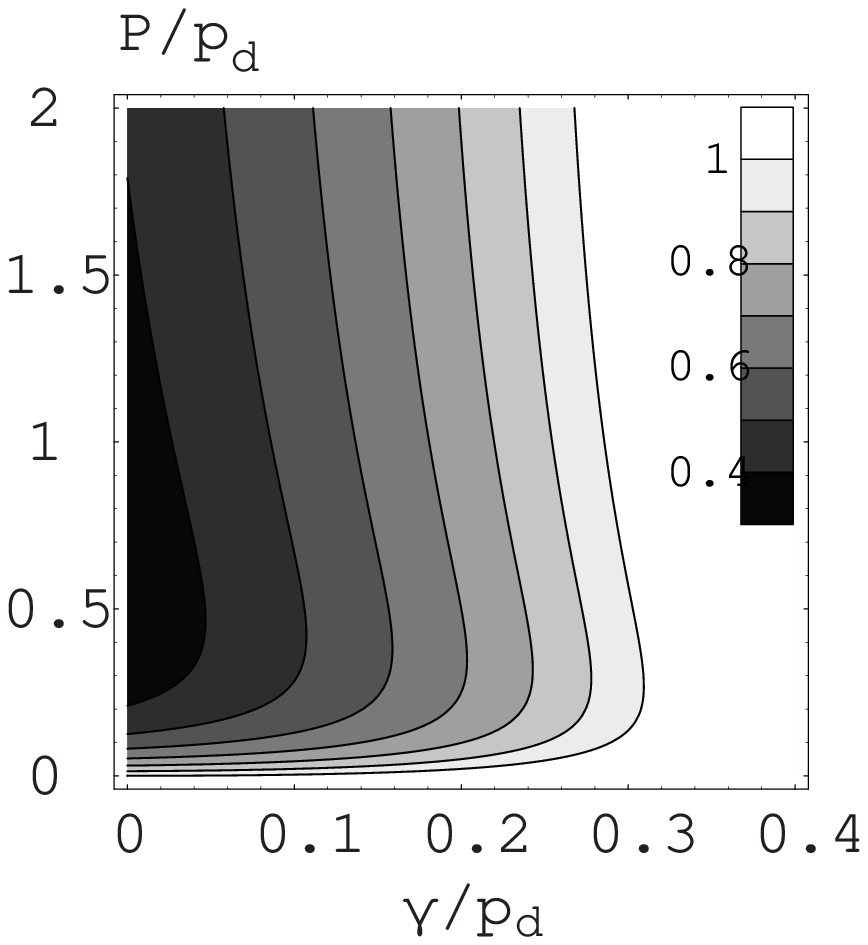} & 
		\includegraphics[width=4.1cm]{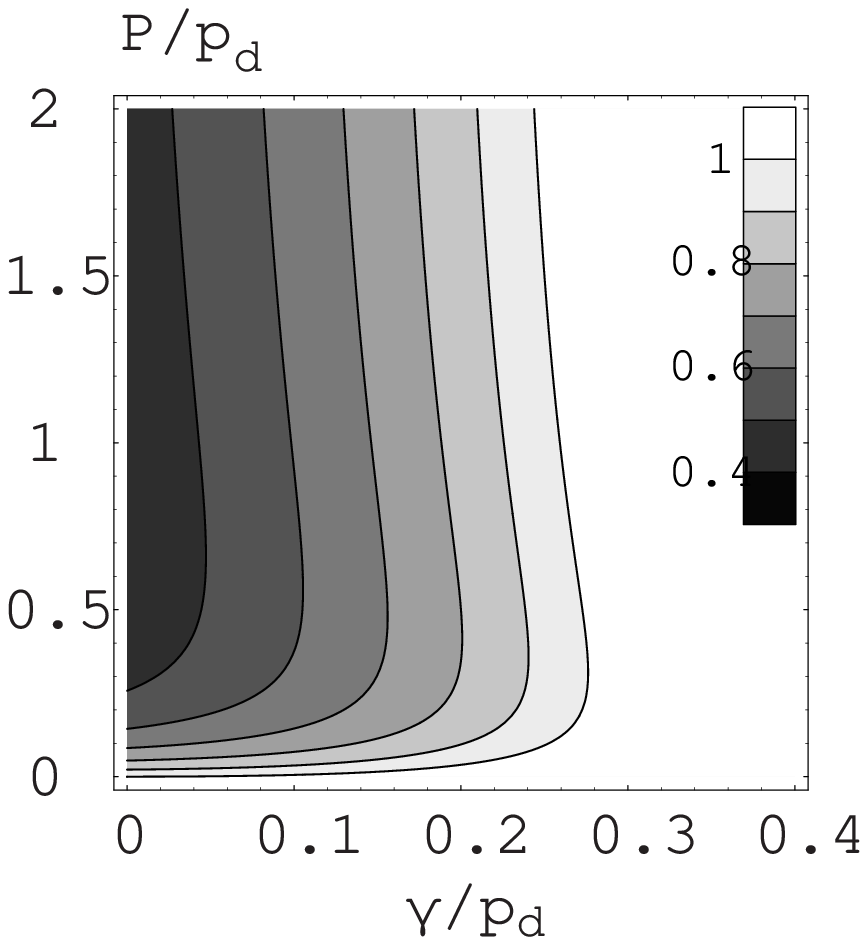} \\
    \end{tabular}
	\caption{Contour plots of the zero-frequency normalized
	photo-current spectral density $\spectral$ for 4-level atoms
	lasers as a function of $\gamma/ p_{d}$ and $P/p_{d}$.  The laser
	parameters are the same as in Fig.~\ref{fig:m}, namely: $N=10^5$,
	$\alpha=6.32$, $p_{d}=632$.  (a) incoherent pumping, $p_{u}=316$,
	(b) coherent pumping, $p_{u}=949$.  In the white area the light
	statistics is super-Poissonian.}
    \label{densities}
\end{figure}

Figure~\ref{densities} gives the normalized zero-frequency
photo-current spectral density $\spectral \left( P/p_{d} ,
\gamma/p_{d} \right)$ in the form of contour plots, selecting $P/p_{u}
= 2 P/p_{d}$ for the case of incoherent pumping and $P/p_{u} =
\frac{2}{3} P/p_{d}$ for the case of coherent pumping.  The darker the
area, the lower is the spectral density.  Since dark areas are wider
in Fig.~\ref{densities}(a) than in Fig.~\ref{densities}(b), incoherent
pumping is to be preferred.  Figure~\ref{densities} shows that
sub-Poissonian light generation by optically pumped 4-level atom
lasers is robust against spontaneous decay and pumping conditions. 
The optimum conditions (darkest areas) are defined in
Eq.~\eqref{SDgam0mincond}.  But small departures from these conditions
do not increase much the noise.

(c) If both $\gamma = 0$ and $N \gg \alpha$, the spectral density
obtained either by setting $\mathscr{N} = \infty$ in
Eq.~\eqref{SDgam0} or $\gamma = 0$ in Eq.~\eqref{SDxnull}, reads
\begin{equation}
   	\label{newS}
	\spectral = 1 - \frac{2 P p_d p_u \left( p_d + 2 \left( P + P \ell
	+ p_u \right) \right) }{{\left( 2 P p_u + p_d \left( P + P \ell +
	p_u \right) \right)}^2} ,
\end{equation}
an expression that coincides with Eq.~(4) of \cite{ritsch:PRA91}.  The
absolute minimum value and corresponding conditions are obtained from 
Eqs.~\eqref{SDgam0min} and \eqref{SDgam0mincond}
\begin{align}
    \label{Min4levels}
	\spectral_{min} &= \frac{1 + 2 \ell}{3 + 4 \ell} , &	
	\frac{P}{p_{u}} &= \frac{1}{1 + \ell} ,
	& \frac{P}{p_{d}} &= \frac{1 + 2 \ell}{2 \left( 1 + \ell \right)} .
\end{align}
For incoherent pumping, $\ell = 0$, we have therefore $\spectral_{min}
= 1/3$ when $p_u = P$ and $ p_d = 2 P$.  For coherent pumping, $\ell =
1$, we have $\spectral_{min} = 3/7$ when $p_u = 2 P$ and $p_{d} =
\frac{4}{3} P$.

Table \ref{optimum} gives the minimum spectral density values
achievable with optically pumped three- and 4-level lasers as
obtained from Eqs.~\eqref{Min4levels}, \eqref{eqB3} and \eqref{eqC3}. 
Under the very special condition of negligible spontaneous decay and
$N\gg \alpha$, the intra-cavity Fano factor depends linearly of the
zero-frequency normalized photo-current spectral density
\cite{golubev:JETP84}, $\spectral = 2 \fano-1$.  Such relation does
not hold in the general situation where Fano factors follow from the
formulation in Sec.~\ref{subsec.Fano}.

\begin{table}
	\caption{Minimum value of the zero-frequency photo-current
	spectral density $\spectral_{min}$ and intra-cavity Fano factor
	$\fano$ for 3 and 4-levels atoms lasers.  The conditions on $P$,
	$p_{u}$ and $p_{d}$ are given.  Spontaneous decay from the upper
	working level is neglected and it is assumed that $N \gg
	\alpha$.}
    \label{optimum}
    \begin{ruledtabular}
		\begin{tabular}{c|c|c|c}
			Laser & $\spectral_{min}$ & $\fano$ & Conditions \\
			\hline
			$\Lambda$--type 3--level\footnotemark[1] & $1/2$ & $3/4$ &
			$p_{d} = 2 P$ \\
			$\Lambda$--type 3--level\footnotemark[2] & $2/3$ & $5/6$ &
			$p_{d} = 3 P$ \\
			$\V$--type 3--level\footnotemark[1] & $1/2$ & $3/4$ &
			$p_{u} = \frac{1}{2} P$ \\
			$\V$--type 3--level\footnotemark[2] & $5/6$ & $11/12$ &
			$p_{u} = \frac{3}{2} P$ \\
			4--level\footnotemark[1] & $1/3$ & $2/3$ & $p_{u} = P$,
			$p_{d} = 2 P$ \\
			4--level\footnotemark[2] & $3/7$ & $5/7$ & $p_{u} = 2 P$,
			$p_{d} = \frac{4}{3} P$ \\
		\end{tabular}
    \end{ruledtabular}
    \footnotetext[1]{Incoherent pumping.}
    \footnotetext[2]{Coherent pumping.}
\end{table}

$\V$--type incoherently-pumped lasers were treated earlier by
\citet{khazanov:PRA90}.  \citet{ralph:PRA91,ralph:QO93} extended the
analysis to incoherently pumped $\Lambda$--type lasers and 4-level
atom lasers.  \citet{ritsch:PRA91} gave a description of 4-level
lasers for the two pumping schemes.  These previous results are
exactly recovered from the present rate-equations method.

Formulas derived in the present paper for coherently-pumped
3-level lasers, however, appear to be new.  Because the present
method involves simple algebra the conditions under which the
photo-current spectral density is minimum are easily obtained.  To our
knowledge, all the results presented in this paper when $\gamma$ does
not vanish, or $N$ is not much larger than $\alpha$, are new.

\subsection{Photo-detection spectrum}
\label{subsec.non0f}

The steady-state relations were given in Eq.~\eqref{steadystateDelta}. 
At some Fourier frequency $\Omega$ the generalized rate equations read
\cite{arnaud:OQE95}
\begin{subequations}
    \label{steadyf}
    \begin{align}
	i \Omega \Delta m & = \Delta R - \Delta Q , \\
	i \Omega \Delta n_{0} & = \Delta D - \Delta J , \\
	i \Omega \Delta n_{1} & = \Delta R + \Delta S - \Delta D , \\
	i \Omega \Delta n_{2} & = \Delta U - \Delta R - \Delta S, \\
	i \Omega \Delta n_{3} & = \Delta J - \Delta U ,
    \end{align}
\end{subequations}

Similar to Sec.~\ref{subsec.0f}, equations Eqs.~\eqref{fluctuations},
\eqref{variations}, \eqref{conservationDelta} and \eqref{steadyf} are
solved for $\Delta Q$.  The formula for the light spectral density
$\spectral$ is the same as Eq.~\eqref{SDfnull} except that the
coefficients $\tilde{c}_{z}$ are complex and frequency dependent
\begin{equation}
	\spectral(\Omega ) = \frac{1}{\alpha m} \sum_{z \in \{j,d,u,r,s\}
	} \tilde{c}_{z} ( \Omega ) \, \tilde{c}_{z}^{\star} ( \Omega ) \,
	\sigma_{z} ,
    \label{SDfnonnull}
\end{equation}
where the Langevin ``forces'' $\sigma_{z}$ are still given in
Eq.~\eqref{SDlangevin}.

After rearranging, Eq.~\eqref{SDfnonnull} gives the spectral density
in the form
\begin{equation}
	\spectral(\Omega ) = 1+ \frac{a_3 \Omega^6 + a_2 \Omega^4 + a_1
	\Omega^2 + a_0}{\Omega^8 + b_3 \Omega^6 + b_2 \Omega^4 + b_1
	\Omega^2 + b_0} ,
    \label{eq:SDOmega}
\end{equation}
where the coefficients $a_i$ and $b_i$ are real.  The form in
Eq.~\eqref{eq:SDOmega} ensures that $\spectral(\Omega )$ tends to
unity (shot-noise level) at high frequencies.

\begin{figure}
    \begin{tabular}{cc}
		(a) & (b) \\
		\includegraphics[width=4.1cm]{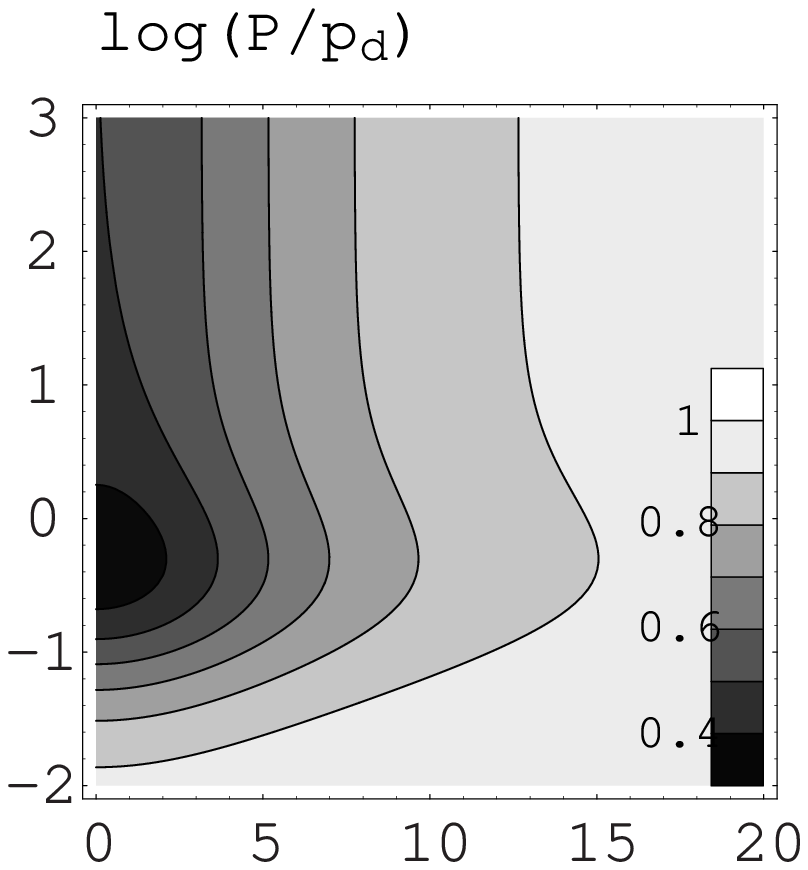} & 
		\includegraphics[width=4.1cm]{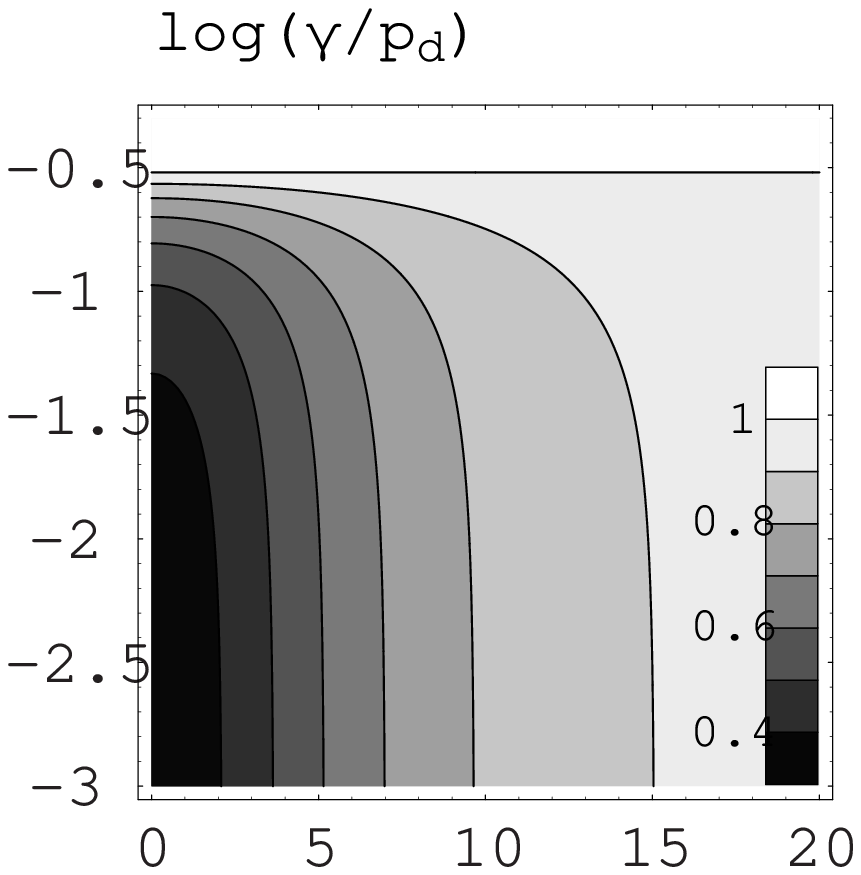} \\
    \end{tabular}
	\caption{Contour plots of normalized photo-current spectra
	$\spectral(\Omega)$ for incoherently pumped 4-level atoms
	lasers.  The laser parameters are the same as in Fig.~\ref{fig:m},
	namely: $N=10^5$, $\alpha=6.32$, $p _{d}=632$, $p_{u}=316$.  (a)
	dependence of $\spectral(\Omega)$ on $P/p_{d}$ with $\gamma =0$. 
	(b) dependence of $\spectral(\Omega)$ on $\gamma /p_{d}$ with $P =
	316$.}
    \label{contours}
\end{figure}

Figure~\ref{contours} shows that $\spectral$ reaches its minimum value
at $\Omega=0$.  When spontaneous decay from the upper working level
may be neglected, light is always sub-Poissonian and the lowest
$\spectral$--value occurs when $P/p_{d}=1/2$.  Spontaneous decay
from the upper working level is inconsequential until $\gamma/p_{d}
\approx 3 \, 10^{-2}$.  The light statistics ceases to be
sub-Poissonian when $P/p_{d} > 0.3$.

\subsection{Fano factor}
\label{subsec.Fano}

The intra-cavity photon statistics is characterized by the Fano factor
$\fano=\ave{\Delta m^2}/\ave{m}$, where $m$ denotes as before the
number of photons in the cavity.  The variance of $m$ equals the
integration over frequency of $\spectral_{\Delta m}(\Omega )$, the
spectral density of $\Delta m$.  Similarly to Sec.~\ref{subsec.non0f},
$\spectral_{\Delta m}(\Omega )$ obtains by solving for $\Delta m$
instead of $\Delta Q$ and its general form is
\begin{equation}
	\spectral_{\Delta m}(\Omega ) = \frac{a'_3 \Omega^6 + a'_2
	\Omega^4 + a'_1 \Omega^2 + a'_0}{\Omega^8 + b'_3 \Omega^6 + b'_2
	\Omega^4 + b'_1 \Omega^2 + b'_0} ,
    \label{eq:SDm}
\end{equation}
where $a'_{i}$ and $b'_{i}$ are real coefficients.
The Fano factor is thus
\begin{equation}
	\fano = \int_{-\infty}^{\infty} \spectral_{\Delta m}(\Omega )
	\frac{\mathrm{d}\Omega}{2 \pi} .
	\label{eq:Fano}
\end{equation}

\begin{figure}
    \includegraphics[width=7.5cm]{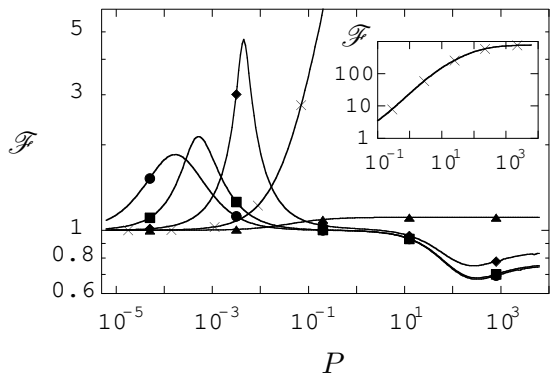} 
	\caption{Internal Fano factor of incoherently pumped 4-level atoms
	lasers as a function of $P$ and $\gamma$; $\bullet$, $\gamma = 0$;
	$\blacksquare$, $\gamma = 6.32 $; $\blacklozenge$, $\gamma = 63.2
	$; $\times$, $\gamma = 632$; $\blacktriangle$, $\gamma = 6325$. 
	The inset is for $\gamma = 632$.  The laser parameters are the
	same as in Fig.~\ref{fig:m}, namely: $N = 10^5$, $\alpha = 6.32$,
	$p_{d} = 632$.}
    \label{fano}
\end{figure}

The Fano factor for incoherently pumped 4-level atoms lasers is
represented as a function of pump and spontaneous decay rates in
Fig.~\ref{fano}.

At small pumping rates, and when the spontaneous decay rate $\gamma$
is large, the photon statistic is essentially that of amplified
thermal light and the Fano factor $\fano = \ave{m}+1$.  The inset in
Fig.~\ref{fano} shows that this simple result holds indeed for
$\gamma=632$, excepts for the highest values of $P$, where a slight
reduction of $\fano$ occurs.

When $\gamma < p_{d}$, the Fano factor exhibits a peak just below
threshold \cite{jones:PRA99}, and decreases to $\approx 1$ until the
ground level gets significantly depleted.  Self-quenching
\cite{mu:PRA92,koganov:PRA00} is not observed here because atomic
polarizations have been adiabatically eliminated.  Otherwise, our
results fully agree with those reported by \citet{koganov:PRA00}.

\section{Conclusion}
\label{sec.Conc}

We have considered optically-pumped 4-level and 3-level atom lasers in
resonant single-mode cavities.  The light statistics has been obtained
from a simple rate-equation approach, using both a Monte Carlo
simulation and an analytical method based on linearization.  The
emitted light may be sub-Poissonian, as was previously observed by
many authors.  Whenever a comparison can be made, exact agreement is
reached with the previous Quantum Optics results.  In the case of
coherently-pumped 3-level atoms lasers our results appear to be new. 
Note that 3-level atoms lasers are not special cases of the 4-level
scheme.  When the assumptions of negligible spontaneous decay and
large atom numbers are not made, the results presented in this paper
for the internal and external field statistics appear to be new.

For practical reasons, Monte Carlo simulations were restricted to $N
\approx 1000$ atoms.  Because the analytical formulas, obtained
through the use of symbolic calculus, are lengthy they were not
written down in the paper.  However, they were employed to determine
the conditions under which the spectral density of the photo-current
reaches its minimum value.  For example, we found that when
spontaneous decay from the upper working level may be neglected,
3-level atoms lasers may deliver light with fluctuations at half the
shot-noise level.  4-level atoms lasers may deliver light with
fluctuations at one third of the shot-noise level.  The photo-current
noise decreases further and tends to zero, under ideal conditions,
when the number of levels becomes large \cite{briegel:PRA96}.

\appendix

\section{ $\Lambda$--type lasers}
\label{sec:App3L}

The number $m$ of photons in the steady-state is still 
\begin{equation}
	m = \frac{1}{2} \left( \mathscr{B} + \sqrt{ \mathscr{B}^2 + 4
	\mathscr{P} \mathscr{N}} \right) , \nonumber
\end{equation}
with
\begin{subequations}
    \begin{align}
	\mathscr{N} &= \frac{N}{\alpha} , \\
	\mathscr{P} &= {\left[ \frac{1}{P} + \frac{2 + \ell}{p_{d}}
	\right]}^{-1} , \\
	\mathscr{B} & = \mathscr{P} \left[ \mathscr{N} - 1 - \ell +
	\frac{1}{p_{d}} \left( \left( 1 + \ell \right) \left(1 + \gamma
	\right) - \mathscr{N} \gamma \right) \right] \nonumber \\
	& \quad - 1 - \gamma .
    \end{align}
\end{subequations}
Lasing may occur if 
\begin{equation}
	\gamma < \frac{\mathscr{N} - 1 - \ell}{\mathscr{N} + 1} p_{d} .
\end{equation}

(a) When spontaneous decay may be neglected, $\gamma = 0$, the
zero-frequency normalized photo-current spectral density reads
\begin{multline}
    \spectral = 1
	+ \frac{2 \left( 1 + \ell + \ell \mathscr{N} \right)}{{\left( 1 +
	\ell - \mathscr{N} \right)}^2}
	+ \frac{4 \left( 2 + \ell \right) \mathscr{P}^2}{p_{d}^2} \\
	- \frac{2 \left( 3 + 2 \ell - 2 \mathscr{N} \right) 
	\mathscr{P}}{\left( 1
	+ \ell - \mathscr{N} \right) p_{d}} .
\end{multline}
Note that when $\ell=0$ this expression 
may be obtained by setting $1/p_{u} = 0$ in the 4-level 
expression. This is not, however, a valid procedure in general.

(b) When $N \gg \alpha$
\begin{multline}
	\spectral = 1
	+ \frac{2 \gamma}{p_{d} - \gamma}
	- \frac{2 \left( 2 \mathscr{P} + \gamma \right)}{p_{d}} \\
	+ \frac{2 \mathscr{P} \left( 2 \left( 2 + \ell \right) \mathscr{P} + 
	\left( 4 + \ell \right) \gamma \right)}{p_{d}^2}
	- \frac{4 \left( 2 + \ell \right) \mathscr{P}^2 \gamma}{p_{d}^3} .
\end{multline}

(c) When $\gamma = 0$ and $N \gg \alpha$
\begin{equation}
	\spectral = 1 - \frac{4 P p_{d}}{{\left( P \left( 2+\ell \right) + 
	p_{d} \right)}^2}
\end{equation}
The minimum value of $\spectral$ and corresponding value of 
$P$ are
\begin{align}
    \label{eqC3}
	\spectral_{min} & = \frac{1 + \ell}{2 + \ell} , &
	P &= \frac{p_{d}}{2 + \ell} .
\end{align}

\section{$\V$--type lasers}
\label{sec:App3V}

The number $m$ of photons in the steady-state is still 
\begin{equation}
	m = \frac{1}{2} \left( \mathscr{B} + \sqrt{ \mathscr{B}^2 + 4
	\mathscr{P} \mathscr{N}} \right) , \nonumber
\end{equation}
with
\begin{subequations}
    \begin{align}
	\mathscr{N} &= \frac{N}{2 \alpha} , \\
	\mathscr{P} &= {\left[ \frac{1}{P} + \frac{1 + 2 \ell}{2 p_{u}}
	\right]}^{-1} , \\
	\mathscr{B} & = \frac{\mathscr{P}}{4 p_{u}} \left[ \left( 2
	\mathscr{N} - 1 \right) \left( \gamma + 2 p_{u} \right) - 1
	\right] \nonumber \\
	& \quad - \frac{1}{2} - \frac{\gamma}{2} - \mathscr{N} \gamma .
    \end{align}
\end{subequations}
Lasing may occur if 
\begin{equation}
	\gamma < \frac{2 \mathscr{N} - 1}{1 + \ell + 2 \ell \mathscr{N} }
	p_{u} .
\end{equation}

(a) When spontaneous decay may be neglected, $\gamma = 0$, the
zero-frequency normalized photo-current spectral density reads
\begin{multline}
    \spectral = 1
	+ \frac{2 \mathscr{N} + 1}{{\left( 2 \mathscr{N} - 1 \right)}^2}
	+ \frac{\left( 1 + 2 \ell \right) \mathscr{P}^2}{2 p_{u}^2} \\
	- \frac{ \left( 4 \mathscr{N} - 1 \right) \mathscr{P}}{2 \left( 2 
	\mathscr{N} - 1 \right) p_{u}} .
\end{multline}

(b) When $N \gg \alpha$
\begin{multline}
	\spectral = 1
	+ \frac{2 \gamma}{\mathscr{P} - \gamma}
	- \frac{\mathscr{P}}{p_{u}}
	+ \frac{\mathscr{P} \left( \mathscr{P} \left( 1 + 2 \ell \right) - 
	\gamma \right) }{2 p_{u}^2} \\
	+ \frac{\left( 1 + 2 \ell \right) \mathscr{P}^2 \gamma}{4 p_{u}^3}
	- \frac{2 \mathscr{P}^2 \gamma}{\left( \mathscr{P} - \gamma
	\right) \left( \mathscr{P} \gamma + 2 \left( \mathscr{P} - \gamma
	\right) p_{u} \right)} .
\end{multline}

(c) When $\gamma = 0$ and $N \gg \alpha$
\begin{equation}
	\spectral = 1 - \frac{4 P p_{u}}{{\left( P \left( 1 + 2 \ell \right) + 
	2 p_{u} \right)}^2}
\end{equation}
The minimum value of $\spectral$ and corresponding value of 
$P$ are
\begin{align}
    \label{eqB3}
	\spectral_{min} & = \frac{1 + 4 \ell}{2 + 4 \ell} , &
	P &= \frac{2 p_{u}}{1 + 2 \ell} .
\end{align}

\begin{acknowledgments}
This work was supported by the STISS Department of Université
Montpellier II and by CNRS under the JemSTIC Program.
\end{acknowledgments}

\bibliography{3levels}

\end{document}